\documentclass[12pt,dvips]{article}

\input{axodraw.sty}
\usepackage{amsmath,amssymb,exscale}
\usepackage{array,multicol}
\usepackage{afterpage,float,flafter}
\usepackage{epsfig,rotating,pifont,fancybox}
\usepackage{cite}
\makeatletter
\@addtoreset{equation}{section}
\makeatother

\title{
\vspace{1cm} 
\Large\textbf{Gauge coupling renormalization in RS1}
\vspace*{.5cm}
\author{\large \textbf{
K.~Agashe\footnote{email: kagashe@pha.jhu.edu} , 
A.~Delgado\footnote{email: adelgado@pha.jhu.edu} 
\mbox{  }and R.~Sundrum\footnote{email: sundrum@pha.jhu.edu}}\\
\emph{
Department of Physics and Astronomy} \\ 
\emph{Johns Hopkins University} \\ 
\emph{3400 North Charles St}. \\ 
\emph{Baltimore, MD 21218-2686}}}

\date{}
\begin{document}
\maketitle
\thispagestyle{empty}
\vspace*{.5cm}
  
\begin{abstract} 
 
We compute the $4D$ low energy effective gauge coupling at one-loop
order in the 
compact Randall-Sundrum scenario with bulk gauge fields and charged 
matter, within controlled approximations. While such computations
are subtle, they can be important for studying phenomenological issues such as
grand unification. Ultraviolet divergences are cut-off using Pauli-Villars
regularization so as to respect $5D$ gauge and general coordinate invariance.
The structure of these divergences on branes and in the bulk is elucidated by
a $5D$ position-space analysis. The remaining finite contributions are obtained
by a careful analysis of the Kaluza-Klein spectrum. 
We comment on the agreement between 
our results and expectations based on the AdS/CFT correspondence, in particular
logarithmic sensitivity to the
$4D$ Planck scale.

\end{abstract} 
  
\newpage 
\renewcommand{\thepage}{\arabic{page}} 
\setcounter{page}{1} 
  
\section{Introduction} 

Extra dimensions provide a theoretically and phenomenologically 
interesting avenue for addressing the Planck-Weak 
hierarchy problem \cite{add}. 
In this paper we focus on the Randall-Sundrum (RS1) 
proposal based on warped compactifications
\cite{rs1}. The geometry of the RS1 vacuum is
a compact slice of AdS$_5$,    
\begin{eqnarray} 
ds^2 & = & e^{-2k |\theta| r_c} \eta_{\mu \nu} dx_{\mu} dx_{\nu} + r_c^2 d 
\theta^2, \; - \pi \leq \theta \leq \pi,
\label{metric}
\end{eqnarray} 
where the extra-dimensional interval is realized as an orbifolded circle of 
radius $r_c$. The two orbifold fixed points, $\theta = 0, \pi$, correspond 
to the ``UV'' (or ``Planck) and ``IR'' (or ``TeV'') branes respectively. In 
warped spacetimes 5D mass scales do not directly correspond to 4D mass 
scales in an effective 4D description, rather the relationship depends on 
location in the extra dimension through the 
 warp factor, $e^{-k |\theta| r_c}$. This allows large 4D mass hierarchies to 
naturally arise without large hierarchies in the defining 5D theory, whose
 mass parameters are taken to be of order the observed 
Planck scale, $M_4 \sim 10^{18}$ GeV.
For example, the 4D massless graviton 
mode is localized near the UV brane \cite{rs1} 
while Higgs physics is taken to be 
localized on the IR brane. In the 4D effective theory one then finds
\begin{equation}
{\rm Weak ~Scale} \sim M_4 e^{-k \pi \langle r_c \rangle}.  
\end{equation}
A modestly large radius can then accommodate a TeV-size weak scale. 
Such a value of the radius
can be 
naturally stabilized using the 
Goldberger-Wise (GW) mechanism \cite{gw1}. 
There is a striking associated  phenomenology of 
TeV-scale Kaluza-Klein (KK) graviton resonances since their wave functions 
are also localized near the IR brane \cite{rs1, hewett}.

The AdS/CFT correspondence 
\cite{adscft}
casts further light on the RS1 mechanism 
\cite{adscftrs, nima1, rattazzi, perez}. 
Effective gravitational field theories in AdS$_5$ encode the constraints of 
{\it four-dimensional} conformal field theory (CFT) 
Ward identities and unitarity, the AdS$_5$ 
loop expansion encodes the constraints of a large-$N$ expansion for the CFT, 
while basic 
CFT data, such as the scaling dimensions of the most relevant operators, are 
encoded in the 5D masses and couplings of the AdS$_5$ theory. This is quite 
analogous to the manner in which chiral lagrangians for pions encode the 
general constraints of chiral symmetry (breaking) and unitarity in a 
transparent and useful way. In terms of the correspondence, the RS1 scenario 
can be viewed as an effective description of a strongly-coupled 4D large-$N$ 
theory coupled to 4D gravity, which is nearly conformal over the Planck-weak 
hierarchy. For phenomenological purposes however, it is more useful 
to employ the AdS$_5$ picture. 

Quantum loops in warped spacetimes are rather subtle because they are 
non-local, and as they span the extra dimension are sensitive to 
greatly varying 4D mass scales. However, their effects can be  important 
to compute, for instance in considering radiative stability of  
radius stabilization \cite{gr1, studentsfuture} 
or in studying grand unification in the RS scenario \cite{pomarol, lisa}.
In a future paper \cite{usfuture}
we will further the exploration of grand unification in 
the RS context. 
In the present paper we study the
loop computation of particular relevance to grand unification, 
namely the one-loop 
correction to the effective 4D gauge coupling when bulk gauge fields  
and bulk charged matter are incorporated into the model 
\cite{pomarol, lisa, choi}.
While this paper was in preparation, reference \cite{gr2} appeared
which has some overlap with our results.
We restrict attention to the effective 4D gauge coupling at energies below 
the lightest KK mass ($\sim$TeV). 
We verify that 
the effective 4D gauge coupling robustly contains a 
logarithmic 
sensitivity to the 4D Planck scale, ``as if'' it had been run down from that 
scale within a purely 4D renormalization group flow \cite{pomarol, lisa}.
All detailed phenomenological issues will
be postponed until reference \cite{usfuture}. For previous  
phenomenological studies of bulk gauge fields in RS1, see references \cite{pheno}.

Several of the results derived here have appeared earlier
\cite{pomarol, lisa}. We have tried to 
give a rather complete treatment here, 
as simple as possible consistent with rigor and independence of 
numerical analysis\footnote{For example, we give a purely 
analytic account of 
the 
Pauli-Villars ``zero-mode'' of reference \cite{pomarol}.}, 
and we have specified the parametric size of corrections 
to our approximations.

The RS1 effective field theory is of course non-renormalizable, or more 
accurately it must be renormalized order by order in the derivative-loop
 expansion. We discuss how this is done in detail in our one-loop calculation. 
We employ a manifestly 5D gauge-invariant and generally covariant UV 
regulator, namely Pauli-Villars, as previously used in reference 
\cite{pomarol}. 
We use a 5D position space analysis to 
straightforwardly understand the structure of the local 
UV divergences on the branes and in the bulk by relating them to the much 
simpler case of flat space compactifications. 
This was also discussed in reference \cite{gr2}.
Renormalization to this order 
is straightforwardly  accomplished. 

The dominant contributions to the finite parts of our calculation are then 
deduced within a mode analysis by careful consideration of the KK 
spectrum.
While our focus is on a particular bulk loop we believe that our 
methodology has broader applicability.

Finally, we discuss the CFT interpretation and compatibility 
of our loop-computation as subleading large-$N$ corrections, in particular, 
the logarithmic 
sensitivity to the 4D Planck scale.

The paper is organized as follows. In section 2 we detail the simple model 
to be studied, namely scalar QED in the bulk. In section 3 we review the 
classical approximation to this model and its CFT interpretation. In 
Section 4 we give a 5D position space analysis of UV divergences and 
renormalization. Section 5 contains a summary of the finite pieces in 
the 4D effective gauge coupling, followed by a detailed derivation using 
KK mode analysis and the results of section 4. Section 6 discusses the CFT 
interpretation of the RS1 loop corrections.
The central results of our paper are contained in Eqs.
(\ref{renorm}), (\ref{0modeloop}), (\ref{resultm50}), 
(\ref{resultm5ltk})
and (\ref{resultm5gtk}).  
 
\section{The Model}
\label{model}

We will consider the simplest model which allows us to focus on
one-loop radiative corrections to the $4D$ effective gauge coupling.  
For this purpose we can fix 
spacetime to be a non-dynamical background of RS1 form, Eq.
(\ref{metric}). The gauge theory 
we consider is 5D scalar QED with a scalar 
mass $m_5$. 
We will study this case in this paper because scalar 
loops are  technically more transparent than loops of higher spin particles. 
Brane-localized charged fields are omitted in this paper; their
loop effects are straightforward to compute.

The dominant part of the action is given by 
\begin{equation} 
S_{\hbox{bulk}} = \frac{1}{g_5^2} \int d^4 x dy \sqrt{-G} \left(
- \frac{1}{4} F_{MN} F^{MN} +  
D_M \phi \left( D^M \phi \right)^{\dagger}
 - m^2_5 |\phi|^2\right), 
\label{Sbulk}
\end{equation} 
where 
$g_5^2$
has dimension (mass)$^{-1}$ and $y=r_c\theta$. 
We will take $A_{\mu}$ and $\phi$ to be 
orbifold-even 
while $A_5$ is taken orbifold-odd.

In addition we can add brane localized terms for our bulk fields,
\begin{equation} 
S_{UV (IR)} = \int d^4 x \sqrt{ -g_{UV(IR)}  } 
\left (- \frac{1}{4} \tau_{UV(IR)}
F_{\mu \nu} F^{\mu \nu} + \sigma_{UV(IR)} \left( D_{\mu} \phi 
\right)^{\dagger} 
D^{\mu} \phi \right ),
\label{Sbrane2}
\end{equation} 
where $\tau_{UV(IR)}, \sigma_{UV(IR)}$ are small dimensionless couplings.  We will 
consider them to be perturbations which we take to be of the same order as 
one-loop processes involving bulk interactions. Thus, working to one-loop order 
the brane-localized vertices do not appear in one-loop graphs, only in 
tree-level graphs. For simplicity, we omit further potential terms for
$\phi$ which would not appear in one-loop graphs.

\section{Review of Classical Approximation}
\label{classical}

We will match our $5D$ model onto a $4D$ effective theory at a scale
provided by the mass of the lightest Kaluza-Klein (KK) excitation.
We will denote this mass scale by 
$m_{KK}$, which is $O(k e^{-k \pi r_c}) \sim O$(TeV) \cite{rs1}. 
Clearly, the scalar does not affect the gauge coupling classically.
The gauge field zero-mode appearing in the $4D$ effective theory is given by
\begin{eqnarray}
A_{\mu} & = & A_{\mu}(x), \nonumber \\
A_5  & = & 0.
\label{0mode}
\end{eqnarray}
Plugging Eq.~(\ref{0mode}) into the action, Eqs.~(\ref{Sbulk}) and 
(\ref{Sbrane2}),
one finds
\begin{equation}
{\cal L}_{eff} \ni -\frac{1}{4 g_4^2} F_{\mu \nu} F^{\mu \nu}, 
\end{equation}
where the effective $4D$ gauge coupling is 
\begin{eqnarray}
\frac{1}{g^2_{ 4 } } &=& \tau_{UV} + \tau_{IR} +  \frac{\pi r_c}{g^2_5}. 
\end{eqnarray}

This coupling can be expressed in a suggestive form, 
\begin{equation}
\frac{1}{g^2_{ 4 } } = \tau_{UV} + \tau_{IR} +  
\frac{\log \big[ O \left( M_4 \right) /{\rm TeV} \big] }{k g^2_5},
\label{0modetree}
\end{equation}
once one puts in 
the RS-GW mechanism \cite{rs1, gw1} 
for generating the Planck-Weak hierarchy without 
fundamental 5D hierarchies:
\begin{equation}
k \pi r_c = \log \big[ \frac{ O \left( M_4 \right) }{ \hbox{TeV} } \big],
\end{equation}
where $M_4$ is the observed $4D$ Planck scale.
The logarithmic dependence on $M_4$ (treated as a variable) 
is a remarkable feature of the RS1 scenario. 
For example, {\em flat} extra dimensions do not automatically possess this 
feature (however see the proposal of \cite{nima2}):
in flat $5D$, we also get $1/ g^2_4 \sim \pi r_c / g^2_5$, but 
unlike in RS1, $r_c$ has no 
relation with the Planck-weak hierarchy.

Eq.~(\ref{0modetree}), though the result of 5D classical physics, 
looks like a quantum gauge coupling in a purely 4D theory which has been 
run down from a Planckian value, $1/g_4^2(\sim {\cal O}(M_4)) \sim 
\tau_{UV}$, with a renormalization group equation, 
\begin{equation}
\mu \frac{d}{d \mu} \frac{1}{g_4^2(\mu)} = - b,
\end{equation}
where $b = 1 / \left( k g_5^2 \right)$ \cite{nima1}.
The running appears to stop at a TeV threshold, with a threshold correction 
$\delta \left( 1/g_4^2 \right)
= \tau_{IR}$. Note that the $\beta$-function coefficient, 
$b$, necessarily has an infrared-free sign and that this 
would happen whether the  bulk gauge field was non-abelian or abelian
(as it is here for the sake of simplicity).
 
These observations are not coincidental, but are in accord with the 
AdS/CFT correspondence \cite{adscft}
applied to the RS1 scenario
\cite{adscftrs, nima1, rattazzi, perez} . The CFT interpretation 
of the RS1 model with bulk gauge field is indeed a purely 4D theory 
consisting of a strongly-coupled large-$N$ 
CFT. The conformal invariance is not 
exact, the central deformations being that a global symmetry of the CFT is 
gauged by a gauge field external to the CFT, 
$A_{\mu}(x)$, and there is a fundamental charged 
scalar field, 
$\phi(x)$, which has a coupling to the CFT at the Planck scale 
of the form, $\delta {\cal L} = 
\phi(x) {\cal O}_{CFT}(x)$, where ${\cal O}_{CFT}(x)$ is an irrelevant
(marginal) 
CFT operator if $m_5 \neq 0$ ($m_5 = 0$). The Planckian value
of the gauge coupling is $1/g_4^2 = \tau_{UV}$.
The CFT is also 
spontaneously broken at the TeV scale.\footnote{Slightly 
irrelevant perturbations at the Planck scale
 can stabilize the resulting Goldstone boson of the 
scale symmetry. This is the dual of the Goldberger-Wise mechanism 
\cite{rattazzi}.}
The leading large-$N$ effects of this dual CFT picture, 
including the running of the 
gauge coupling of $A_{\mu}$, are captured by classical effects in the RS1 
picture. In particular Eq.~(\ref{0modetree}) describes the running, valid between the Planck 
and TeV scales, due to CFT charged matter. Despite the fact that the CFT 
itself is strongly self-coupled  (although not strongly coupled to the 
external gauge field), the running can be 
summarized by a single constant, $b$, as is familiar in one-loop 
perturbation theory. Here, this fact follows directly from the conformal 
invariance of the charged matter dynamics. The fact that $b$ has infrared-free
sign (whether or not $A_{\mu}$ is Abelian or non-Abelian, so long as the rank
of the gauge group is fixed in the large-$N$ limit) follows from the 
fact that the charged matter comes in complete large-$N$ representations which 
always overwhelm asymptotically free contributions to $b$. At the TeV 
threshold where scale invariance is broken, 
there is a threshold correction to the gauge coupling $\delta \left( 1/g_4^2 \right) = 
\tau_{IR}$. 

Note that although the CFT interpretation of Eq.~(\ref{0modetree}) is the
gauge coupling at momenta $q \ll$ TeV, the $\log q$ dependence one would 
expect from $\phi$ loops (if the $\phi$ mass is tuned to be 
sufficiently small) is absent. This is 
because $\phi$ loops are subleading in the large-$N$ 
expansion, 
while the classical 
approximation in the RS1 scenario corresponds to leading order. 
That is, in large-$N$ the external gauge coupling is scaled to be of 
order $1/\sqrt{N}$ so that running only arises above $\sim$TeV where this 
suppression is compensated by  large-$N$ multiplicities. 
Below $\sim$TeV the dual 
of our RS1 theory has at most a single charged scalar and so there is no 
multiplicity enhancement. We will see the running effects due to  
$\phi$ loops
when we include subleading large-$N$ effects, corresponding to loop effects in 
our model. We now turn to these RS1 loop effects.

\section{Quantum Divergence Structure and Renormalization}
\label{divergence}

The most straightforward way to understand the divergence structure of the 
one-loop vacuum polarization is to view the Feynman diagrams 
in position space. They can be formally expressed in terms of 
the $\phi$-propagator, $G(x, y; x', y')$. It satisfies the defining equation,
\begin{equation}
\square_{(x,y)} G(x, y; x', y') = \delta^4 (x- x') \delta(y - y'), 
\end{equation}
where $\square$ is the d'Alembertian in the 5D warped background, subject 
to the orbifold boundary conditions, 
\begin{equation}
\partial_y G_{|boundaries} = 0.
\end{equation}

Recall that we are taking brane-localized terms to be negligible within
one-loop diagrams. The Feynman diagrams can only 
diverge when the initial and final points in a 
propagator coincide in the  Feynman position-integral. Thus all divergences 
must be local and must correspond to either 
local bulk terms or local brane terms. To determine what the relevant 
local divergence structures are it is useful to use the fact that
for short-distance 
propagation in the vicinity of divergences
the finite AdS radius of curvature is irrelevant and therefore
the propagators can be approximated by their flat space equivalents.
For bulk divergences we need only consider the 5D gauge theory in 5D 
Minkowski spacetime, while for brane-localized divergences we need only 
consider orbifolds of 5D Minkowski spacetime. Once the flat space divergences
are determined their warped equivalents are obtained by inserting the metric
dependence using general covariance 
(also see the discussion in reference \cite{gr2}).
It is simpler to 
euclideanize the whole problem and replace 5D Minkowski space by 5D Euclidean 
space, where the bulk propagator (away from any orbifolds) is 
$1/|X-X'|^3$ for short distances. 

\begin{figure}[t]
\centering
\begin{picture}(300,100)(0,0)
\Photon(0,50)(50,50){5}{4}
\Vertex(50,50){2}
\DashCArc(75,50)(25,0,180){3}
\DashCArc(75,50)(25,180,360){3}
\Vertex(100,50){2}
\Photon(100,50)(150,50){5}{4}
\Text(75,15)[]{(A)}
\Photon(175,50)(275,50){5}{8}
\Vertex(225,50){2}
\DashCArc(225,75)(25,0,360){3}
\Text(225,15)[]{(B)}
\end{picture}
\caption{One-loop diagrams contributing to the vacuum polarization}
\label{diagfig}
\end{figure}
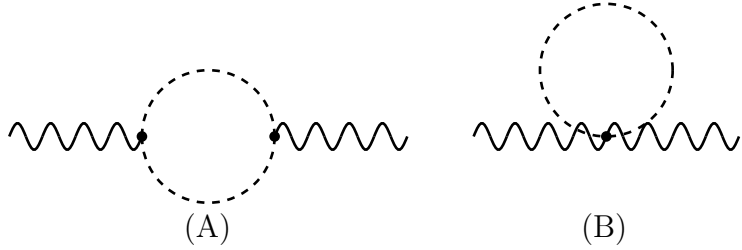

The one-loop vacuum polarization is given by the sum of the two 
diagrams in Fig.~\ref{diagfig}. 
We will first examine potential bulk divergences by 
considering the contributions of spacetime vertices away from orbifolds. 
Fig.~\ref{diagfig}B contains a singular factor 
$G(x, y; x, y)$. We can write this as a limit of point-splitting,
 Limit$_{x', y' \rightarrow x,y}
G(x, y; x', y')$, whereupon we see that the diagram is cubically 
divergent. It is also straightforward to see that as the two vertices in 
Fig.~\ref{diagfig}A approach each other in the Feynman integral, 
this diagram is also 
cubically divergent. We can regulate the divergences gauge-invariantly using 
the standard method of Pauli-Villars (PV) regularization, with a set of scalar 
regulator fields $\psi$, some with fermionic statistics,
 having the same form of action as $\phi$, 
\begin{equation} 
S_{\hbox{bulk}} = \int d^4 x dy \sqrt{-G} \left(
D^M \psi \left( D_M \psi \right)^{\dagger}
-\Lambda^2_5 |\psi|^2\right). 
\end{equation} 
This cut-off represents the unknown physics due to bulk states with large 5D 
masses which have been integrated out to yield our model as an effective 
field theory. We will consider $\Lambda_5 \gg k, m_5$. 
We must understand the extent to which our results 
are sensitive to this unknown physics. By standard power-counting  it is 
clear that the local cubic bulk divergences in Fig.~\ref{diagfig} multiply 
$A_M^2$. Since our regulator is gauge-invariant such divergences 
cancel between the two diagrams as usual. Power-counting 
and gauge invariance 
now shows that there 
remains a single linear divergence in Fig.~\ref{diagfig} multiplying 
$F_{MN}^2$. 

Now let us consider what divergences emerge when the Feynman vertices 
approach an orbifold fixed hyperplane \cite{georgi, gw2}. 
We can write the propagator between two 
points $X, X'$ for 
an orbifolded 5D Euclidean space 
as a superposition of a pure Euclidean space 
propagator for those two points and a pure Euclidean space propagator from 
the $Z_2$ image of $X$ in the orbifold ``mirror'' to $X'$,
\begin{equation}
G(X, X') = \frac{1}{2} \frac{1}{|X - X'|^{3}} +  
 \frac{1}{2} \frac{1}{|X_P - X'|^{3}},
\end{equation}
where $X_P$ is just the ``mirror image'' of $X$ in the orbifold fixed 
hyperplane. We then find that beyond the expected bulk divergence discussed 
above, which persists in the vicinity of the brane, Fig.~\ref{diagfig} gives 
rise to a new type of divergence from cross terms involving the 
product (of derivatives) of $1/|X - X'|^{3}$ with 
$1/|X_P - X'|^{3}$. By power-counting, these can only lead to 
logarithmic or quadratic divergences in 
the Feynman integral when both $X$ and $X'$ coincide on the orbifold 
fixed hyperplane. The Pauli-Villars regularization cuts off these divergences 
as well and gauge invariance again forces the quadratic divergences 
multiplying $A_{\mu}^2 \delta(y)$ to vanish, leaving only a logarithmic 
divergence multiplying $F_{\mu \nu}^2 \delta(y)$. 

Technically, cubic and quadratic 
divergences require three regulators fields. However, since
these divergences are gauge non-invariant and cancel we need only use a single
PV field to regulate the remaining gauge invariant linear and logarithmic 
divergences. We will do this from now on.

It is now simple to translate the 5D orbifolded Euclidean space divergence 
structures to the RS1 spacetime. To this we can add the finite parts of 
the effective action for the gauge field up to one-loop order. The result is
\begin{eqnarray}
\Gamma &=&  \int d^4 x dy \sqrt{-G} 
c\Lambda F_{MN} F^{MN} + \int d^4x \sqrt{-g_{UV}} \left(
-\frac{b_4 \log \Lambda}{4} F_{\mu \nu} F^{\mu \nu} \right) \nonumber \\ 
 & & + \int d^4x \sqrt{-g_{IR}} \left(
-\frac{b_4 \log \Lambda}{4} F_{\mu \nu} F^{\mu \nu} \right) 
+ \hbox{finite non-local one-loop corrections}.
\label{effaction}
\end{eqnarray}
We have not computed the precise coefficient $c$ of the linear divergence 
because it is highly regularization-scheme dependent. In any scheme
it is a number of $\mathcal{O} (1/24 \pi^3)$, a 5D loop factor. 
Brane-localized logarithmic  
divergences are however not regularization dependent 
since the logarithms must contain finite mass/energy scales in 
order to  balance dimensions. Thus the coefficients must be physical and we
have written the results in terms of the $\beta$-function coefficient 
for purely 4D massless scalar QED, $b_4 = 1/ \left( 24 \pi^2 \right)$. 

Eq.~(\ref{effaction}) demonstrates that all one-loop cut-off 
sensitivity can be  eliminated 
by renormalization of couplings we have already included, 
\begin{eqnarray}
\frac{1}{ g_{5~R}^2(k) } &\equiv & c(\Lambda - k)+ \frac{1}{g_5^2}, 
\nonumber \\
\tau_{UV(IR)~R}(k) &\equiv& \frac{b_4}{4} \log \frac{ \Lambda }{k} + 
\tau_{UV(IR)}.
\label{renorm}
\end{eqnarray}

The only sensitivity to the unknown massive 
physics represented by the cut-off is parameterized by these renormalized 
values.

In {\em flat} $5D$ space, 
a similar linear $\Lambda$ dependent correction to     
$1/ g^2_5$  is 
usually referred to as ``power-law running'' \cite{ddg}. In contrast, in RS1, 
this leads to
$\delta \left( 1 / g^2_{4} \right) 
\sim  c\Lambda\pi r_c \sim 
c \Lambda / k \log \big[M_4/\hbox{TeV} \big]$, which appears as a one-loop
Planckian logarithm;
this logarithm has the same origin as in the 
case of the tree-level coupling.

We can now write a general form for 
the one-loop corrected gauge coupling in the effective 4D 
theory below $m_{KK}$: 
\begin{equation}
\frac{1}{g_4^2(q)} 
= \tau_{UV~R}(k) + \tau_{IR~R}(k) +  \frac{\pi r_c}{g^2_{5~R}(k)} 
+ f(q, r_c, m_5, k), 
\label{0modeloop}
\end{equation}
where $f$ is finite as $\Lambda \rightarrow \infty$. This limit is a 
reasonable approximation since we are considering $\Lambda \gg k, m_5$. 

We now turn to calculating the dominant behavior of $f$. Since it is finite, 
we can no longer use 5D locality and the simplifications of relating 
AdS$_5$ locally to 5D Minkowski space. Therefore it is no longer 
profitable to use a position space analysis. Instead we make use of the 
KK decomposition of all states to exploit the preserved 4D Poincare 
symmetry of the RS1 background.

\section{KK mode analysis} 
\label{kk} 

\subsection{Set-up}
 
 
To calculate the renormalized gauge coupling in the 4D effective 
theory at one-loop order, 
we consider the sum of $4D$
vacuum polarization diagrams with  
charged (physical + PV) KK states in the loop
and the zero-mode of the gauge field on external legs (with $q \ll$ TeV)
\cite{pomarol}. 
The couplings in each diagram are completely fixed 
by 4D gauge invariance, and the signs of each diagram fixed by the 
Bose (Fermi) statistics of the physical (PV) charged fields. Thus, only the 
4D mass spectrum is needed to compute the diagrams. There are two potential 
sources of UV divergence. Each 4D diagram has standard divergences, while 
there can be a further divergence in the sum over KK towers. However, 
these
divergences must be cut off by the 
PV regularization as seen in section \ref{divergence}.
The basic idea is to pair up each physical mode contribution with 
a PV mode contribution: 
this pair gives a finite
one-loop contribution depending only on the masses of the modes.
The fact that PV 
regularization provides a complete cut-off in position space then implies that
the sum over all such pairs also converges.

We first summarize the results for three regimes of $m_5$:
$m_5 \ll q \ll m_{KK} \sim O$(TeV), $m_{KK} \ll m_5 \ll k$ and $m_5 \gg k$.
We then derive these results by combining a detailed 
mode analysis 
with our earlier study of UV divergences in section \ref{divergence}.
While the results are very simple in form (for example, the terms sensitive
to $r_c$ are linear), they do not follow entirely from simple general
considerations. Our detailed mode analysis appears necessary for proof.

\subsection{Summary of results}

For $m_5 \ll q \ll m_{KK} \sim O$(TeV), we get (cf.~Eq.~(\ref{0modeloop}))
\begin{eqnarray}
f(q,r_c,m_5,k)
& = & b_4 \left( \log \frac{k}{q} +  \xi k \pi r_c
+ O(1) \right),
\label{resultm50}
\end{eqnarray}
where $\xi$ is a constant of order one.
Note that the $\xi$-dependent effect can be renormalized away by a
straightforward modification of first line of Eq.~(\ref{renorm}).
By ``$O(1)$''  we refer to terms which are insensitive to any of our formal
large parameters, for example these terms are bounded as $\Lambda / k 
\rightarrow 
\infty$
or $k \pi r_c \rightarrow \infty$.

For $\hbox{TeV} \ll m_5 \ll k$,
we get
\begin{eqnarray}
f(q,r_c,m_5,k)
& = & 
b_4 \left( 
\log \frac{k}{m_5} - \frac{m_5^2}{8 k} \pi r_c+  \xi k \pi r_c + O(1) \right).
\label{resultm5ltk}
\end{eqnarray}
Note that the ``$\xi$''
appearing in Eqs.~(\ref{resultm50}) and (\ref{resultm5ltk}) are the same, 
but the $O(1)$ terms are different.

This case 
will be very important 
when we study GUTs because we will encounter $X$, $Y$ 
gauge bosons with
bulk masses $m_5 \stackrel{<}{\sim} k$ \cite{usfuture}.
Reference \cite{pomarol} considered
the case where the PV mass $\Lambda \ll k$ and found the same dependence on 
$\Lambda$ as the 
$m_5$ dependence
in Eq.~(\ref{resultm5ltk}). 
The relevance of such a mass scale smaller than $k$
for grand unification, although in a 
different scenario than \cite{usfuture}, 
was also pointed out in \cite{pomarol}.


The case $m_5 \gg k$ is straightforward. By reasoning along the same 
lines as in section 
\ref{divergence} for the cut-off dependence, we get
\begin{eqnarray}
f(q,r_c,m_5,k)
& = & \frac{b_4}{2} \log \frac{k}{m_5} + c \left(k - m_5  \right) \pi r_c.
\label{resultm5gtk}
\end{eqnarray}
We will not discuss this case any further.


\vspace{0.2cm}

\subsection{Derivation of results}
\label{kkdetails}
We first need to discuss the KK spectrum for physical and PV charged modes. 

For a 
scalar with $5D$ mass $m_5$, 
the classical wave equation of motion (neglecting $\sigma_{UV, IR}$ terms in 
Eq.~(\ref{Sbrane2}) 
at this order)
determines the spectrum of 
KK mass eigenvalues \cite{gw3, gp}: 
\begin{equation} 
b_{\nu} \left( \frac{m_n}{k} \right) 
= b_{\nu} \left( \frac{ m_n }{k} \; e^{k \pi r_c} \right), 
\label{eigeneqn} 
\end{equation} 
where 
\begin{equation} 
b_{\nu} 
(x)
= \frac{ (2 - \nu) J_{\nu} 
(x)
+ 
x
J_{\nu - 1} 
(x)
} 
{ (2 - \nu) Y_{\nu} 
(x)
+ 
x
Y_{\nu - 1} 
(x)
} 
\end{equation} 
and $\nu = \sqrt{ 4 + m_5^2 / k^2 }$.
For the PV KK spectrum, $\Lambda$ replaces $m_5$.
 
\subsubsection{$m_5 \ll q \ll m_{KK}$}

\vspace{0.2cm}

{\it \large The spectrum}

\vspace{0.2cm}

There are four distinct regions of $m_n$ for physical and PV fields: (a) 
$m_n \stackrel{<}{\sim} \Lambda \; e^{- k \pi r_c}$,
(b) $\Lambda \; e^{-k \pi r_c} \ll m_n \ll k$,
(c) $k \stackrel{<}{\sim}
m_n \stackrel{<}{\sim} \Lambda$ and
(d) $m_n \gg \Lambda$.
In the following, $\nu \approx 2$ for physical modes and 
$\nu = \sqrt{4 + \Lambda ^2 / k^2 }$ for PV modes. 

(a) $m_n \stackrel{<}{\sim} \Lambda \; e^{- k \pi r_c}$

\vspace{0.2cm}

In this region, because $e^{k \pi r_c} \gg 1$, we get $m_n \ll k$.
So, the 
number of eigenvalues in this region
can only depend on 
$\Lambda / k$
and not on
$k \pi r_c$, because the LHS of Eq.~(\ref{eigeneqn}) is approximately zero. 
We denote this number by $N_{(a)}^{phys}$ and
$N_{(a)}^{PV}$, respectively.

It is straightforward to check that there is a single mode 
with 4D mass $\ll m_{KK}$, where the arguments of both LHS and RHS of 
Eq.~(\ref{eigeneqn}) are small. This mode is the lightest physical state 
(which is a zero mode for $m_5=0$). For $0<m_5\ll m_{KK}$ its mass is 
\begin{equation}
m_0 \approx \frac{m_5}{\sqrt{2}}.
\label{lightest}
\end{equation}

\vspace{0.2cm}

(b) $\Lambda\; e^{-k \pi r_c} \ll m_n 
\ll k$  
 
\vspace{0.2cm} 
 
Here, 
the argument of the LHS of Eq.~(\ref{eigeneqn})
$\ll 1$, where we use the approximation 
\begin{equation}
b_{\nu} (x)
\stackrel{x \ll 1}{\approx}
\frac{ 2 
\left( x / 2 \right) ^{2 \nu} \left( 1 + 2 / \nu \right) 
( 1 - \nu ) \; \sin \nu \pi \; \Gamma ( 1 - \nu )}
{ \Gamma ( \nu ) \; \big[  x^2 - 2 (2-\nu) (1-\nu) \big] }.
\label{discon}
\end{equation}
The RHS of Eq.~(\ref{eigeneqn}) can be approximated by 
$\cot  
\left[ m_n / k \; e^{k \pi r_c} - 
\pi / 2 \;\nu  +\pi / 4 + O \left( k \; e^{-k \pi r_c} / m_n 
\right) \right]$, where the error term is
$\nu$-dependent.
The KK mass eigenvalues are therefore given by \cite{gp}
%
\begin{equation} 
m_n = \left( n - \frac{3}{4} + \frac{1}{2} \nu + 
O \left( n^2 \; e^{-2 k \pi r_c} \right)
+ O \! \left( \frac{1}{n} \right) \right) \pi k 
e^{-k \pi r_c}, 
\label{mnlessthank} 
\end{equation} 
where the error terms are 
$\nu$-dependent. Note that in this region $n \; e^{-k \pi r_c} \ll 1$. 

(c) $k \stackrel{<}{\sim} 
m_n \stackrel{<}{\sim}\Lambda$

\vspace{0.2cm}

Here, the LHS of Eq.~(\ref{eigeneqn}) is a piecewise smooth function, 
i.e., it has discontinuities at zeroes of the denominator of
$b_{\nu} ( m_n / k )$, but is otherwise smooth. Since it is independent of
$k \pi r_c$, the number of discontinuities in each tower in this region
depends only on $\Lambda / k$. 
The RHS of Eq.~(\ref{eigeneqn}) is approximated $\cot  
\left[ m_n / k \; e^{k \pi r_c} - 
\pi / 2 \;\nu  +\pi / 4 \right]$ as in (b).

Let us divide the $4D$ mass spectrum in this region into ``bins'' of
size equal to the period of the RHS
of Eq.~(\ref{eigeneqn}), i.e., $\pi k e^{-k \pi r_c}$. We see that
each tower has one eigenvalue per bin, except possibly at discontinuities of
the LHS,
where there may be zero or two eigenvalues in a bin.
We will refer to bins with such a discontinuity 
(for either physical or PV case) 
as 
``exceptional  bins'', and the modes in these bins as ``exceptional modes''.
Thus, the number of exceptional modes, denoted  
by 
$N_{(c)}^{phys}$ and
$N_{(c)}^{PV}$, respectively, depend on $\Lambda / k$ and not 
on $k \pi r_c$.
Note that this
one-eigenvalue-per-bin rule holds in region (b), with no exceptions, by 
Eq.~(\ref{mnlessthank}).

(d) $m_n \gg \Lambda$

\vspace{0.2cm}

Here, 
$b_{\nu} \left( m_n / k \right) 
\approx \cot \left( m_n / k - \pi / 2 \;  \nu + \pi/4
\right)$ while 
$b_{\nu} \left( m_n / k \; e^{k \pi r_c} \right)$ is approximately \\
$\cot  
\left[ m_n / k \; e^{k \pi r_c} - 
\pi / 2 \; \nu  +\pi / 4 \right]$. 
Thus, the eigenvalues are given by 
%
\begin{equation}
m_n \approx \frac{k \pi n}{e^{k \pi r_c} - 1}. 
\label{mnmorethank}
\end{equation}
The corrections to Eq.~(\ref{mnmorethank}) are
$O(1/n)$ for large $n$ and are $\nu$-dependent.

\vspace{0.2cm}

{\it \large Pairing modes}

\vspace{0.2cm}

Having discussed the mass spectrum in the four regions of $m_n$, 
we now turn to
the question of how to pair one-loop contributions
of physical and PV modes in each region.
From our analysis in section \ref{divergence}, we know that
the one-loop contribution summed over PV and physical modes in
{\em all} regions is UV finite.

In region (d), we pair the one-loop
contributions of physical
and PV modes with the same $n$. Using
Eq.~(\ref{mnmorethank}) and approximating the sum over 
modes as an integral, we get a {\em finite}
and $r_c$-{\em in}dependent contribution 
to $f$ (cf. Eq.~(\ref{0modeloop})) from the {\em infinite} number of
modes in region (d).
The error in this approximation is $O(e^{-k \pi r_c})$.

Thus, the one-loop contribution 
from the remaining {\em finite} number of (physical + PV) modes in the 
regions (a), (b) and (c) must also be UV finite, i.e.,
the logarithmic divergences have to cancel between
physical and PV diagrams. This is possible if 
and only if
the number  
of modes in the regions (a)-(c) is the same for the two towers.
Since the one-eigenvalue-per-bin rule is valid in region (b), this region
has the same number of physical and PV modes
which can be paired up. Also,
the modes in non-exceptional bins 
in region (c) can be paired up.
Thus, the sum of the number
of modes in region (a) and the number of exceptional modes in region
(c) has to be the same for physical and PV towers, i.e.,
$N_{(a)}^{phys} + N_{(c)}^{phys} = N_{(a)}^{PV}
+ N_{(c)}^{PV}$. 

In region (b), we pair the contribution of physical and PV modes in 
the {\em same} bin. 
Using Eq.~(\ref{mnlessthank}) and approximating the sum over these bins 
as an
integral gives
\begin{eqnarray}
\delta_{1-\hbox{loop}} \left( 1/ g^2_4 (q) \right) 
& \stackrel{\Lambda \gg k}{\approx} 
& b_4 \left( k \pi r_c 
\; \hbox{Frac} \! \left( \frac{\Lambda}{2k} - 1 \right) + 
s \! \left( \frac{\Lambda}{k}, k \pi r_c \right) \right), 
\label{kkm501}
\end{eqnarray}
where
$\hbox{Frac} (x) $ 
denotes the fractional part of $x$ and $s$ 
can be bounded by a function of $\Lambda / k$ alone.


In region (c), 
we do not have a good analytic approximation for the spectrum other than
the one-eigenvalue-per-bin rule in 
non-exceptional bins. Using this rule, 
the sum over these bins 
can be easily bounded:
\begin{eqnarray}
\delta_{1-\hbox{loop}} \left( 1/ g^2_4 (q) \right) 
& < & b_4 \left( \log \frac{\Lambda}{k} + O(1) 
\right).
\label{kkm502}
\end{eqnarray}
%

Finally, we consider exceptional modes 
in region (c) and {\em all} modes in region (a) together. 
The number of physical
modes is
$N_{(a)}^{phys} + N_{(c)}^{phys}$ and number of PV modes
is $N_{(a)}^{PV} + N_{(c)}^{PV}$.
Recall that they
are equal and we can pair them arbitrarily. If both 
modes in a pair come from the same region, the one-loop logarithm is
$r_c$ independent. If not, the one-loop logarithm is $\propto r_c$.
The lightest physical mode
(the zero-mode as $m_5 \rightarrow 0$) is an exception to this rule 
because, unlike other modes in region (a), $m_0 \ll q \ll m_{KK}$ 
(see Eq.~(\ref{lightest})).
Therefore, we get
\begin{eqnarray}
\delta_{1-\hbox{loop}} \left( 1/ g^2_4 (q) \right) & \approx & 
b_4 \left( k \pi r_c \; u \left( \Lambda / k \right) 
+ v \left( \Lambda /k \right) + \log k / q \right),
\label{kkm503}
\end{eqnarray}
where $u$, $v$ are $r_c$ independent because the
$N$'s are independent of $r_c$. We have dropped power-suppressed 
$q$-dependence in all contributions. Similarly, we have dropped 
power suppressed $m_5$ dependence since $m_5 \ll q$. 
For the lightest mode, $q$ dependence is 
non-analytic and is therefore retained. This 
$\log q$ dependence is 
entirely determined within the low energy
$4D$ effective theory where we have a 
single light charged scalar.

Putting together contributions from all regions and using the $\Lambda$ 
sensitivity already determined in Eqs.~(\ref{renorm}) and (\ref{0modeloop}),
Eq.~(\ref{resultm50}) follows.

\subsubsection{$\hbox{TeV} \ll m_5 \ll k$}

\vspace{0.2cm}

{\it \large The spectrum}

\vspace{0.2cm}

The PV spectrum is as before. 

The physical spectrum in the four regions is as follows 
(in the following $\nu = \sqrt{4 + m_5^2 / k^2} \approx 2 
+ m_5^2 / \left( 4 k^2 \right)$).

(a) 
$m_n \stackrel{<}{\sim} \Lambda \; e^{- k \pi r_c}$

\vspace{0.2cm}

The eigenvalues are shifted relative to those in the case $m_5 \ll$TeV 
by $\sim m_5^2 /k \; e^{-k \pi r_c} \ll k \; e^{-k \pi r_c} \sim
\hbox{spacing between eigenvalues}$,\footnote{The eigenvalues
in the case $m_5 \ll$TeV are given by
$k \; e^{-k \pi r_c} \times \hbox{zeroes of} \; J_1 (x)$.} except
that there is no analog of the light mode 
(cf. Eq.~(\ref{lightest})) in this region.
Thus, 
\begin{eqnarray}
N_{(a)}^{phys} \left( \hbox{TeV} \ll m_5 \ll k \right)
& = & N_{(a)}^{phys} \left( 
m_5 \ll \hbox{TeV} \right) - 1,
\nonumber \\
N_{(a)}^{PV} \left( \hbox{TeV} \ll m_5 \ll k \right) & = &
N_{(a)}^{PV} \left( 
m_5 \ll \hbox{TeV} \right).
\label{Nrelation}
\end{eqnarray}

(b) $\Lambda \; e^{-k \pi r_c} \ll m_n \ll k$

\vspace{0.2cm}

Previously, for $m_5 \ll$TeV, 
there were no exceptional bins in this region. Now, however,
the LHS of Eq.~(\ref{eigeneqn})
has a single discontinuity as can be seen from Eq.~(\ref{discon})
%
%
and the RHS of Eq.~(\ref{eigeneqn})
is approximately
$\cot  
\left[ m_n / k \; e^{k \pi r_c} - 
\pi / 2 \; \nu + \pi / 4 
+ O \left( k \; e^{-k \pi r_c} / m_n 
\right) \right]$. 
This discontinuity gives rise to a single
exceptional physical
mode near $m_5 / \sqrt{2}$.\footnote{This is analogous to the
PV ``zero-mode'' which plays a central role in the analysis of
\cite{pomarol}.}
Far away from the discontinuity, $m_n \ll m_5$ or $m_n \gg m_5$,
Eq.~(\ref{mnlessthank}) continues to hold. For $m_n \sim O(m_5)$, 
other than the exceptional mode, all we will use is that the 
one-eigenvalue-per-bin rule holds.

(c) $k \stackrel{<}{\sim} 
m_n \stackrel{<}{\sim} \Lambda$ 

\vspace{0.2cm}

As before, the modes follow the one-eigenvalue-per-bin rule with some 
$N_{(c)}^{phy}$, $N_{(c)}^{PV}$ ``exceptions''.

(d) $m_n \gg \Lambda$

\vspace{0.2cm} 

The spectrum (Eq.~(\ref{mnmorethank})) and analysis
of corrections is as before.

\vspace{0.2cm}

{\it \large Pairing modes}

\vspace{0.2cm}

The pairing
in region (d) follows just as before. 
The sum over non-exceptional modes 
in region (b) is given by
\begin{eqnarray}
\delta_{1-\hbox{loop}} \left( 1/ g^2_4 (q) \right) & \approx &
b_4 \left( k \pi r_c \big[ 
\hbox{Frac} \! \left( \frac{\Lambda}{2k} - 1 \right) - 
\frac{m_5^2}{ 8 k^2 } \big] + 
t \! \left( \frac{\Lambda}{k}, \frac{m_5}{k},
k \pi r_c \right) \right),
\end{eqnarray}
where $t$ is bounded by a function of $\Lambda / k$ alone.
Although Eq.~(\ref{mnlessthank}) does not hold for $m_n \sim O(m_5)$
as discussed above, the contribution of these modes can easily be bounded 
and shown to affect only $t$.
The bound on the contribution
of non-exceptional modes in region (c) is unchanged from
the $m_5 \ll$TeV case (Eq.~(\ref{kkm502})).

Now consider 
modes in region (a) and exceptional modes in regions (c) and (b).
The number of physical and PV modes are equal. We will pair 
the exceptional
physical mode in region (b) with an 
exceptional PV mode in region (c). All the modes in region (a) and the
remaining exceptional modes
in region (c) can be paired arbitrarily. All these pairings yield
\begin{eqnarray}
\delta_{1-\hbox{loop}} \left( 1/ g^2_4 (q) \right) & \approx & 
b_4 \left( k \pi r_c \; u \left( \Lambda / k \right) 
+ w \left( \Lambda /k \right) + \log k / m_5 \right).
\label{kkm5ltk3}
\end{eqnarray}
Note that because of Eq.~(\ref{Nrelation}), $u$ appearing
in Eqs.~(\ref{kkm5ltk3}) and (\ref{kkm503}) is the same. 
The contribution of 
the lightest mode for 
$m_5 \ll q \ll$TeV is 
replaced by the 
contribution of the exceptional mode in region (b). The functions 
$w$ and $v$ in 
Eqs.~(\ref{kkm5ltk3}) and (\ref{kkm503})
need not be equal. Using the $\Lambda$ sensitivity determined in
Eqs.~(\ref{renorm}) and (\ref{0modeloop}) and above observations,
Eq.~(\ref{resultm5ltk}) follows.

\section{The dual CFT: subleading corrections}
\label{cft}

As discussed earlier, 
the RS model in classical approximation is dual to a $4D$ CFT picture
at leading order in a large-$N$ expansion. The one-loop RS corrections
are then dual to sub-leading large-$N$ corrections \cite{adscft}.  
Our one-loop RS results, Eqs.~(\ref{renorm}), (\ref{0modeloop}),
(\ref{resultm50}) and (\ref{resultm5ltk}) can be reexpressed in the dual form:
\begin{eqnarray}
\frac{1}{g_4^2 (q)} &=& \tilde{\tau}_{UV} + \tilde{b}  
\log \big[ \frac{ O \left( M_4 \right) }{ \hbox{TeV} } \big] 
+ \tilde{\tau} _{IR} + b_4 \log \frac{ \hbox{TeV} }{q}, \;
\hbox{for} \; m_5 \ll q \nonumber \\
 &=& \hat{\tau}_{UV} + \hat{b} \log \big[ 
\frac{ O \left(M_4 \right) }{ \hbox{TeV} } \big] 
+ \hat{\tau} _{IR}, \; \hbox{for} \; m_5 \gg \hbox{TeV}.
\end{eqnarray}
In the dual interpretation, $\tilde{\tau} _{UV}$ or $\hat{\tau}_{UV}$ 
sets the Planckian gauge
couplings, the CFT charged matter and $\phi (x)$ lead to logarithmic
running down to the TeV threshold with $\beta$-function coefficients
$\tilde{b}$ or $\hat{b}$, $\tilde{\tau} _{IR}$ 
or $\hat{\tau}_{IR}$ is a TeV-threshold correction and 
$b_4 \log \hbox{TeV} / q$ is the running 
due to $\phi$ in the case where its mass is much smaller than
$q$. All of the coefficients receive sub-leading corrections
(compared to the classical approximation/leading order in large-$N$
expansion in section \ref{classical})
except for $b_4$ which vanishes at leading-order.
It is not yet known how to compute
the detailed form of these corrections, for example, the $m_5$ dependence
(in the case $m_5 \gg$ TeV) directly from CFT considerations, but the RS
picture allows us to estimate them, as we have done in section \ref{kk}.

Note that it is significant that 
the loop
corrections in the RS model admit the 
dual interpretation. For example, if Eqs.~(\ref{resultm50}) and
(\ref{resultm5ltk}) had contained
a contribution $\propto \sqrt{k \pi r_c} \sim \sqrt{\log \big[ 
O \left( M_4 \right) / TeV \big] }$, the
CFT interpretation would fail.  

 
 

\section*{Acknowledgments} 

We thank Lisa Randall, Matthew Schwartz and Neal Weiner for useful 
discussions. K.~A.~ is supported by the Leon Madansky fellowship and by
NSF Grant P420D3620414350. A.~D.~ is supported by NSF Grants P420D3620414350 
and
P420D3620434350.
R.~S.~ is supported by
a DOE Outstanding Junior Investigator award Grant P442D3620444350 and 
NSF Grant P420D3620434350.

\end{document}